\documentstyle[12pt]{article}

\begin{document}

{\sf \begin{center}
\noindent
{\Large \bf Analytical solutions of neutrino wave equations in Kerr geometry with Vaidya-Patel coordinates}\\[3mm]

by \\[0.3cm]

{\sl L.C. Garcia de Andrade}\\ 

\vspace{0.5cm}
Departamento de F\'{\i}sica
Te\'orica -- IF -- Universidade do Estado do Rio de Janeiro-UERJ\\[-3mm] 
Rua S\~ao Francisco Xavier, 524\\[-3mm]
Cep 20550-003, Maracan\~a, Rio de Janeiro, RJ, Brasil\\[-3mm]
Electronic mail address: garcia@dft.if.uerj.br\\[-3mm]

\vspace{2cm}
{\bf Abstract}
\end{center}
\paragraph*{}

Analytical solution of Weyl neutrino wave equation in Kerr geometry is presented by making use of the two-spinor component spin-coefficient Newman-Penrose (NP) calculus. So far only asymptotic or approximate solutions have been found for the Weyl equation in this background. It is shown that neutrino current asymmetry is also present in this solution.
\vspace{0.5cm}
\noindent
{\bf PACS numbers:} \hfill\parbox[t]{13.5cm}{0420, 0450} 

\newpage

\paragraph*{}

Earlier Chandrasekhar \cite{1} has shown that the Weyl neutrino wave equation in Kerr geometrical background can be separated in oblate coordinates and an approximate solution has been found. Although Chandrasekhar has made use of the NP calculus as well we show here that he failed in finding a analytic solution for this problem because of an inapropriate choice of coordinates. As an astrophysicist he used the well known Boyer-Lindquist coordinates \cite{2} suitable for black holes investigations. Here we show that the use of the powerful mathematical tool of NP calculus allied with the Vaidya-Patel \cite{3} null coordinates help us to find an analytical solution for the problem using the simple and well known method of the separation of variables. This problem seems to be important for General Relativity (GR) and astrophysics since can avoid sometimes the use of complicated approximate methods such as numerical calculus , although this method is useful in many realistic situations. With our solution no need is made to solve the Teukolsky \cite{4} equation for the specific case of spin-$\frac{1}{2}$ particles. We also show that neutrino asymmetric currents \cite{5} are present in this solution. The neutrino equations 
\begin{equation}
{\Delta}{\psi}=({\gamma}- {\pi}){\psi}
\end{equation}
\begin{equation}
\bar{{\delta}}{\psi}= ({\alpha}- {\mu}){\phi}
\end{equation}
Here ${\psi}$ is the neutrino field obtained by making use of neutrino spinor
\begin{equation}
{\psi}^{A}=({\psi}) {i^{A}}
\end{equation}
where $i^{A}$ is a spinor basis dyad. 
From this tetrad we can obtain the following  differential operators
\begin{equation}
{\delta}=-\frac{\bar{\rho}}{\sqrt{2}}(iasinx\frac{{\partial}}{{\partial}u}+\frac{{\partial}}{{\partial}x}+icosecx\frac{{\partial}}{{\partial}y})
\end{equation}
\begin{equation}
{\Delta}={\bar{\rho}}{\rho}[-\Upsilon\frac{{\partial}}{{\partial}r}+{\Omega}\frac{{\partial}}{{\partial}u}+a\frac{{\partial}}{{\partial}y}]
\end{equation} 
where we made use of the variables
\begin{equation}
{\Omega}=r^{2}+a^{2}
\end{equation}
and 
\begin{equation}
\Upsilon=\frac{r^{2}+a^{2}-2m r}{2}
\end{equation}
Note that here the neutrino field ${\psi}$ depends on the null coordinate $u$ since although the Kerr metric \cite{6} does not depend on u the neutrino is a wave and does depend on this coordinate. The spin coefficients appearing in the Weyl equation are given by 
\begin{equation}
{\alpha}-{\pi}= -\bar{\beta}=\frac{cotx{\rho}}{2\sqrt{2}}
\end{equation}
and
\begin{equation}
{\gamma}-{\mu}=[r-m]\frac{{\rho}\bar{\rho}}{\sqrt{2}}
\end{equation}
Substitution of these operators into the Weyl equations yields the following system of PDE
\begin{equation}
ia sinx\frac{{\partial}{\psi}}{{\partial}u}-\frac{{\partial}{\psi}}{{\partial}x}+icscx\frac{{\partial}{\psi}}{{\partial}y}={cotx{\rho}}{\psi}
\end{equation}
\begin{equation}
-{\Upsilon}\frac{{\partial}{\psi}}{{\partial}r}+{\Omega}\frac{{\partial}{\psi}}{{\partial}u}+ a\frac{{\partial}{\psi}}{{\partial}y}= \frac{[r-m]}{\sqrt{2}}{\psi}
\end{equation}
To obtain the solution let us make the following ansatz 
\begin{equation}
{\psi}(u,r,x,y)= e^{i{{\Lambda}u}}R(r,x,y)
\end{equation}
where c is a constant. This ansatz represents a plane wave representation. This ansatz reduces the partial differential equation to 
\begin{equation}
(acsinxR+\frac{cotx}{2})R+\frac{{\partial}{R}}{{\partial}x}=icscx\frac{{\partial}{R}}{{\partial}y}
\end{equation}
\begin{equation}
{\Upsilon}\frac{{\partial}{R}}{{\partial}r}-(ic{\Omega}+m -r)R=a\frac{{\partial}{R}}{{\partial}y}
\end{equation}
Let us now try to further reduce this system by making the separation of variables
\begin{equation}
R(r,x,y)= A(r)B(x)C(y)
\end{equation}
Substitution of this expression into  the PDE yields
\begin{equation}
\frac{{\Upsilon}}{A}\frac{d{A}}{dr}-(ic{\Omega}+m -r)=a\frac{dC}{dy}=c_{0}
\end{equation}
\begin{equation}
(c_{0}cscx-acsinx-\frac{cotx}{2})=\frac{1}{B}\frac{dB}{dx}
\end{equation}
where here $c_{0}$ represents the separation variables constant. These equations can be immeadiatly integrated and collecting them we finally obtain the neutrino wave function in Kerr geometry 
\begin{equation}
{\psi}_{Kerr}(u,r,x,y)= \frac{1}{2}\frac{e^{acosx-i[c_{0}y+c((r-u)+ln{\Upsilon}^{m}+(2m^{2}-ac_{0}){\Gamma})]}}{{\Upsilon}sin^{\frac{1}{2}}}
\end{equation}
where  ${\Gamma}=\frac{arctan\frac{(r-m)}{\sqrt{\Theta}}}{\sqrt{\Theta}}$ where ${\Theta}=a^{2}-m^{2}$. By considering the neutrino current defined as
\begin{equation}
J^{\mu}= {\psi}\bar{\psi} l^{\mu}
\end{equation}
where $l^{\mu}$ is a tetrad leg of the tetrad frame $(l^{\mu},n^{\mu},m^{\mu},\bar{m^{\mu}})$. one obtains the following neutrino current in Kerr background
\begin{equation}
J^{\mu}= {\Upsilon}^{-2}\frac{e^{a^{2}cos^{2}x}}{4sinx} l^{\mu}
\end{equation}
where to simplify the expression we consider $c_{0}=0$. One notices from this current that the neutrino asymmetry \cite{7} is present in the Kerr background since the signe of the angular momentum density of the black hole $a=\frac{J}{m}$ changes the neutrino current. Since the Einstein-Weyl equation is known to be equivalent to the Einstein-Cartan-Weyl \cite{8,9} ,the solution presented here can be useful in constructing solutions in spacetimes with torsion. In this case the neutrino current could be expressed in terms of the Cartan contortion as $K_{{\alpha}{\beta}{\lambda}}=k{\epsilon}_{{\alpha}{\beta}{\lambda}{\theta}}J^{\theta}$. To resume we have shown that a analytical solution of neutrino wave equations in Kerr geometry is possible by a proper choice of null coordinates used by Vaidya and Patel in the construction of Kerr radiative spacetimes, since this solution reduces to the Kerr solution in the constant mass case. 

\section*{Acknowledgements}
\paragraph*{}
I am very much indebt to Professors I.D. Soares and P.S.Letelier for helpful discussions on the subject of this paper. Special thanks go to Professor Jerry Griffiths for his kind attention and interest in our work. Grants from CNPq (Ministry of Science of Brazilian Government) and Universidade do Estado do Rio de Janeiro (UERJ) are acknowledged.


\begin{thebibliography}{9}
\bibitem{1} S. Chandrasekhar, The Mathematical Theory of Black Holes, Oxford University Press (1983).
\bibitem{2} C.P. Boyer and Lindquist, Phys. Rev. D 
\bibitem{3} P.C. Vaidya and L. K. Patel, Phys. Rev. D 7(1973) 3590.
\bibitem{4} M. Carmeli, General Relativity and Gauge Fields (1982) JP and S. Teukolsky and S. Shapiro, Black Holes, White Dwarfs and Neutron stars: The physics of Compact stellar systems (1983).
\bibitem{5} I.D. Soares, Phys. Rev. D (1978), I.D. Soares, PhD Thesis (CBPF) (1976),in portuguese and I. D. Soares,Phys. Rev. D23,272,2(1981) and I.D. Soares and L. M. Rodrigues, Phys. Rev. D (1985).
\bibitem{6} R. P. Kerr, Phys. Rev. Lett. (1963).
\bibitem{7} A. Vilenkin, Phys. Rev. D 20,8 (1979) 1807.
\bibitem{8} P. S. Letelier,Phys. Lett. A (1975) 331.
\bibitem{9} J. B. Griffiths, Phys. Lett. 75 A (1980)441 and S. Jogia and P. Griffiths, General Relativity and Gravitation (1980).
\end{thebibliography}
\end{document}